\documentclass[a4paper,fleqn,usenatbib]{mnras}

\usepackage{newtxtext,newtxmath}

\usepackage[T1]{fontenc}
\usepackage{ae,aecompl}

\usepackage{graphicx}	
\usepackage{amsmath}	
\usepackage{amssymb}	
\usepackage{breakurl}

\newcommand{\etal}{{et al}\/.}

\title[FR-II radio galaxies at low frequencies II]{FR-II radio galaxies at low frequencies II: spectral ageing and source dynamics}

\author[J.J.~Harwood \etal]{Jeremy J.\ Harwood$^{1}$\thanks{E-mail: jeremy.harwood@physics.org}, Martin J.\ Hardcastle$^{2}$, Raffaella Morganti$^{1,3}$,
\newauthor Judith H. Croston$^{4}$, Marcus Br\"uggen$^{5}$, Gianfranco Brunetti$^{6}$,
\newauthor Huub J. A. R\"ottgering$^{7}$, Aleksander Shulevski$^{1}$, Glenn J. White$^{4,8}$
\\$^{1}$ASTRON, The Netherlands Institute for Radio Astronomy, Postbus 2, 7990 AA, Dwingeloo, The Netherlands
\\$^{2}$Centre for Astrophysics Research, School of Physics, Astronomy and Mathematics, University of Hertfordshire, College Lane,\\ Hatfield, Hertfordshire AL10 9AB, UK
\\$^{3}$Kapteyn Astronomical Institute, University of Groningen, P.O. Box 800, 9700 AV, Groningen, The Netherlands
\\$^{4}$School for Physical Sciences, The Open University, Milton Keynes MK7 6AA, England
\\$^{5}$Universit\"at Hamburg, Hamburger Sternwarte, Gojenbergsweg 112, 21029 Hamburg, Germany 
\\$^{6}$INAF- Istituto di Radioastronomia, via P. Gobetti 101, 40129 Bologna, Italy 
\\$^{7}$Leiden Observatory, Leiden University, P.O. Box 9513, 2300 RA Leiden, The Netherlands 
\\$^{8}$RAL Space, The Rutherford Appleton Laboratory, Chilton, Didcot, Oxfordshire OX11 0NL, England
}

\date{Accepted XXX. Received YYY; in original form ZZZ}

\pubyear{2016}

\begin{document}
\label{firstpage}
\pagerange{\pageref{firstpage}--\pageref{lastpage}}
\maketitle

\graphicspath{{./images/}}

\begin{abstract}

In this paper, the second in a series investigating FR II radio galaxies at low frequencies, we use LOFAR and VLA observations between 117 and 456 MHz in addition to archival data to determine the dynamics and energetics of two radio galaxies, 3C452 and 3C223, through fitting of spectral ageing models on small spatial scales. We provide improved measurements for the physical extent of the two sources, including a previously unknown low surface brightness extension to the northern lobe of 3C223, and revised energetics based on these values. We find spectral ages of $77.05^{+9.22}_{-8.74}$ and $84.96^{+15.02}_{-13.83}$ Myr for 3C452 and 3C223 respectively suggesting a characteristic advance speed for the lobes of around one per cent the speed of light. For 3C452 we show that, even for a magnetic field strength not assumed to be in equipartition, a disparity of factor of approximately 2 exists between the spectral age and that determined from a dynamical standpoint. We confirm that the injection index of both sources (as derived from the lobe emission) remains steeper than classically assumed values even when considered on well resolved scales at low frequencies, but find an unexpected sharp discontinuity between the spectrum of the hotspots and the surrounding lobe emission. We suggest that this discrepancy is due to the absorption of hotspot emission and/or non-homogeneous and additional acceleration mechanisms and, as such, hotspots should not be used in the determination of the underlying initial electron energy distribution.

\end{abstract}


\begin{table*}
\centering
\caption{List of target sources and galaxy properties}
\label{jvlatargets}
\begin{tabular}{llcccccccc}
\hline
\hline
Name&IAU Name&Redshift&178 MHz&5 GHz Core&Spectral index&LAS&Size&Revised&Revised\\
&&&Flux (Jy)&Flux (mJy)&(178 to 750 MHz)&('')&(kpc)&LAS ('')&Size (kpc)\\
\hline
3C223&J0936$+$361&0.137&9.0&16.0&0.74&306&740&329&796\\
3C452&J2243$+$394&0.081&130&59.3&0.78&280&428&287&438\\
\hline
\end{tabular}
\vskip 5pt
\begin{minipage}{16.0cm}
`Name' and `IAU Name' list the 3C and IAU names of the galaxies. `Spectral Index' lists the low frequency spectral index between 178 to 750 MHz, `LAS' the largest angular size of the source and `Size' its largest physical size. The `Redshift', `178 MHz Flux', `5 GHz Core Flux', `Spectral Index' , `LAS' and `Size' column values are taken directly from the 3CRR database \citep{laing83} (\url{http://3crr.extragalactic.info/cgi/database}). `Revised' values of the source sizes are listed in the last 2 columns and are measured from the 368 MHz P-band VLA images presented in this paper. 
\end{minipage}
\end{table*}

\section{Introduction}
\label{intro}

\subsection{Radio galaxies}
\label{rgintro}

Radio loud active galaxies, while only constituting a small fraction of the population as a whole, have over the previous decades become an increasingly important part in our understanding of how galaxies evolve to form the universe we observe today. Models of galaxy evolution have historically been unable to unify simulations with observations giving an excess of cold gas, a deficit in hot gas, and an overestimation of the star formation rate (e.g. \citealp{balogh01}). In order to correct these discrepancies, it is generally agreed that an additional energy input is required in order to expel cold gas from galaxies and increase the cooling timescales of hot gas if one is to correctly determine the evolutionary path of galactic populations (e.g. \citealp{cowie77, fabian77}). Radio galaxies, highly energetic structures which are observed on scales of up to a few Mpcs in size (e.g. \citealp{bridle76, leahy91, mullin06, marscher11}), are therefore one of the currently favoured mechanisms by which energy can be transferred to the surrounding environment in a self regulating manner (e.g. \citealp{croton06, bower06, fabian12, mcnamara12, morganti13, heckman14}). Understanding what drives these powerful radio sources, the timescales on which they operate, and the effect they have on their external environment is therefore crucial in our understanding of galaxy evolution as a whole.

Due to their steep radio spectra ($\alpha \geq 0.5$)\footnote{We define the spectral index such that $S \propto \nu^{-\alpha}$}, the energetics of the lobes of FR II radio galaxies are dominated by the low energy electron population. Recent studies have suggested that this population may have a significantly higher total energy content (the combined particle and magnetic field energies) than has previously been assumed when considered using modern analytical methods (\citealp{harwood13, harwood15, harwood16}; herein H13, H15 and H16), potentially impacting on our understanding of not only the energetics of FR IIs but also on their age and dynamics. However, such studies have yet been able to directly probe the low-frequency spectrum on well resolved scales. The limited sensitivity, image fidelity, and frequency coverage of the previous generation of radio interferometers has historically restricted earlier studies of FR IIs to $\sim\,$1 arcsec resolution observations at GHz frequencies and low resolution observations at metre wavelengths, leaving well resolved studies of FR IIs at low frequencies a relatively unexplored region of parameter space. New instruments such as the LOw Frequency ARray (LOFAR; \citealp{haarlem13}) and the recently upgraded Karl G. Jansky Very Large Array (VLA) are now able to resolve this issue by providing the high quality, high resolution observations at MHz frequencies required to undertake studies that are critical to our understanding of this population.

\begin{figure*}
\centering
\hspace{-0.5cm}\includegraphics[angle=0,height=8.5cm]{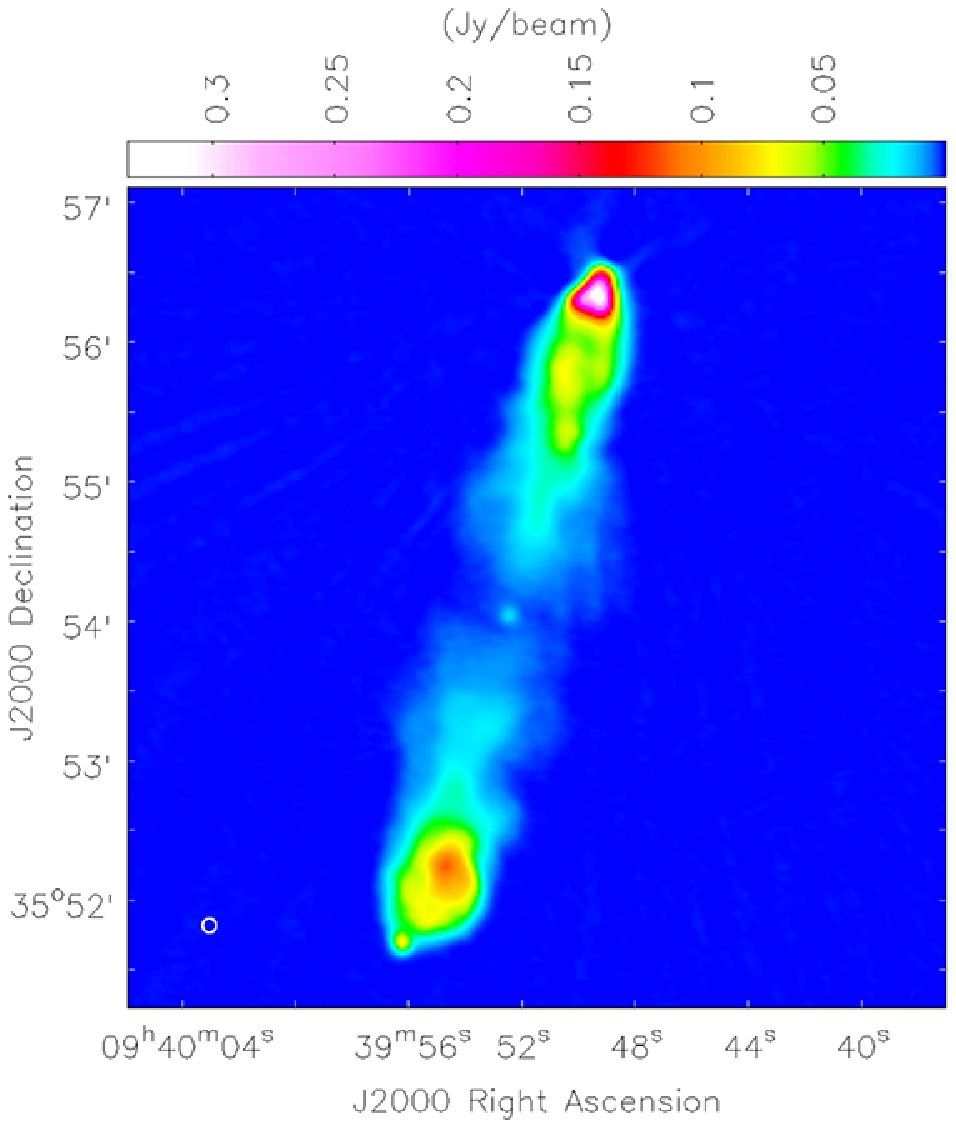}\hspace{1.5cm}
\includegraphics[angle=0,height=8.5cm]{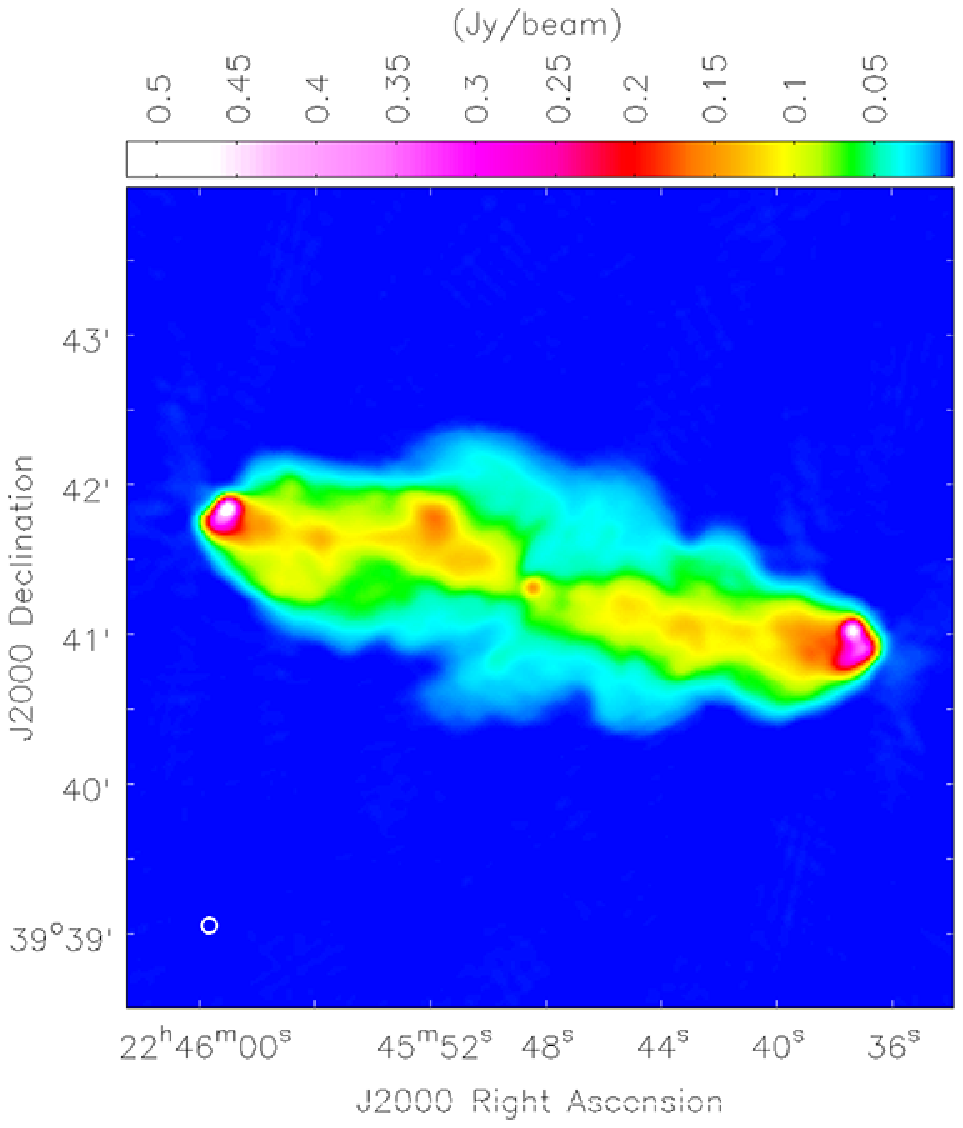}\\
\caption{Full bandwidth radio maps of 3C223 (left) and 3C452 (right) at 368 MHz, imaged in \textsc{casa} using multiscale CLEAN and \emph{nterms} = 2. The off-source RMS noise is $0.16$ mJy beam$^{-1}$ for 3C223 and $0.17$ mJy beam$^{-1}$ for 3C452. The restoring beam of 6 arcsec is indicated in the bottom left corner of the images.}
\label{pbandimages}
\end{figure*}

\subsection{Spectral ageing in FR-II radio galaxies}
\label{specageintro}

One of the key values required for deriving the properties of radio galaxies, such as their total power output and the speed at which their hotspots are advancing through the external medium, is the duration for which the source has been in its active phase. As the lobes of FR II radio galaxies radiate via the synchrotron process, in theory, for an electron population with a power law energy distribution that is initially described by \begin{equation}\label{initialdist} N(E) = N_0E^{-\delta} \end{equation} and knowing that in a fixed magnetic field a single electron will have energy losses, $\tau$, that scale as a function of frequency such that \begin{equation}\label{freqscale} \tau = \frac{E}{{\rm{d}} E / {\rm{d}} t} \propto \frac{1}{E} \propto \frac{1}{\nu^{2}} \end{equation} at later times (assuming that no further particle acceleration is present) the observed population can be described by \begin{equation}\label{poploss} N(E,\theta,t) = N_{0} E^{-\delta} (1 - E_{T} E)^{-\delta-2}\end{equation} where $E_{T}$ are model dependent losses which are a function of the pitch angle of the electrons to the magnetic field, $\theta$, and the time since they were accelerated, $t$. Through modelling of the resulting curvature, known as spectral ageing, one is able to determine the time since acceleration of a given region of plasma, hence the age of the source as a whole.

These single injection spectral ageing models and their application have recently been discussed in detail by H13 and H15 and are one of the most commonly used observationally driven methods for determining the age of radio galaxies (e.g. \citealp{jamrozy05, kharb08, degasperin12}); however, it has recently been shown that many of the assumptions which go in to these models are not as robust as had been previously thought. The general reliability of these models in determining the intrinsic age of a source has long been a matter of debate (e.g. \citealp{alexander87, eilek96, eilek97, blundell00}) and although it has been shown that through a combination of revised parameters (H13; H15; \citealp{turner16}; H16) and more complex ageing models (e.g. Tribble model, \citealp{tribble93, hardcastle13a, harwood13}) many of these issues can be resolved, new questions have come to light and assumptions must still be made about the low frequency spectrum on small spatial scales. 

One of the key questions raised by recent studies is the correct value of the power law index for the low energy electron distribution described by Equation \ref{initialdist} used within models of spectral ageing. At the point of acceleration, $\delta$ is assumed to be directly related to the observable spectrum of the plasma with an index (known as the injection index) given by\footnote{At near minimum energy for first order Fermi acceleration the spectrum is not a power law in momentum which further complicates this relation (e.g. \citealp{brunetti02, amato06}). We use here the classical assumption but note this effect likely needs addressing in future models.}.  \begin{equation}\label{alphainject}\alpha_{inj} = \frac{\delta - 1}{2}\end{equation} This parameter not only has a significant impact on the determination of the energetics of a source (e.g. \citealp{croston04, croston05}; H16) but also on it derived age, hence the dynamics. Based on theoretical arguments (e.g. \citealp{blandford78}) and observations of hotspots (e.g. \citealp{meisenheimer89, carilli91}) values of around $\alpha_{inj} = 0.5$ to $0.6$ have traditionally been adopted but, through derivation of the injection index at GHz frequencies (H13, H15), it has been suggested that it may be significantly steeper than previously thought taking values of around $0.7$ to $0.8$. Such steep spectral indices have previously been seen in observations of the integrated flux of FR IIs (e.g. the 3CRR sample, \citealp{laing83}) but were generally attributed to curvature downstream due to ageing in the oldest regions of plasma. However, as was shown by H16, their spectrum can remain steep down to very low frequencies ($\approx\,$10 MHz) where such ageing effects should be negligible. Determining the well resolved, low frequency spectrum of these sources is therefore key if we are to reliably determine the injection index and understand the particle acceleration processes occurring in FR IIs, a vital step in advancing our understanding of radio galaxy physics and galaxy evolution as a whole.

In this paper, the second in a series examining FR II radio galaxies at low frequencies, we use LOFAR and VLA observations to explore the spectrum, dynamics and particle acceleration processes of two powerful radio sources on well resolved scales and will address 4 primary questions:

\begin{enumerate}
\item What is the spectrum of FR II radio galaxies on small spatial scales at low frequencies?\\
\item What are the model parameters and spectral age of our sample when tightly constrained at low frequencies?\\
\item How do these observational findings compare to dynamical models of FR IIs?\\
\item How does this impact on our understanding of the age, dynamics, and energetics of FR IIs?
\end{enumerate}

In Section \ref{method} we give details of target selection, data reduction and the analysis undertaken. Section \ref{results} presents our results and in Section \ref{discussion} we discuss these findings in the context of the aims outlined above. Throughout this paper, a concordance model in which $H_0=71$ km s$^{-1}$ Mpc$^{-1}$, $\Omega _m =0.27$ and $\Omega _\Lambda =0.73$ is used \citep{spergel03}.

\begin{table*}
\centering
\caption{Summary of images by frequency}
\label{imagedetails}
\begin{tabular}{lcccccccccc}
\hline
\hline
Source&Frequency&Off-source RMS&\multicolumn{6}{c}{Integrated flux (Jy)}&Reference\\
&(MHz)&(mJy beam$^{-1}$)&Lobe 1&$\pm$&Lobe 2&$\pm$&Total&$\pm$&\\
\hline

3C223	&	118.4	&	1.08	&	13.69	&	1.37	&	11.00	&	1.10	&	24.69	&	2.47	&	1	\\
	&	125.4	&	0.947	&	12.89	&	1.29	&	10.28	&	1.03	&	23.17	&	2.32	&	1	\\
	&	132.5	&	0.910	&	12.03	&	1.20	&	9.76	&	0.98	&	21.79	&	2.18	&	1	\\
	&	146.5	&	0.784	&	10.88	&	1.09	&	8.85	&	0.89	&	19.65	&	1.97	&	1	\\
	&	153.6	&	0.840	&	10.38	&	1.04	&	8.40	&	0.84	&	18.78	&	1.88	&	1	\\
	&	160.6	&	0.834	&	9.63	&	0.96	&	7.84	&	0.78	&	17.47	&	1.75	&	1	\\
	&	280.0	&	0.462	&	6.34	&	0.32	&	4.89	&	0.24	&	11.23	&	0.56	&	2	\\
	&	296.0	&	0.311	&	6.33	&	0.32	&	4.88	&	0.24	&	11.21	&	0.56	&	2	\\
	&	312.0	&	0.274	&	5.97	&	0.30	&	4.67	&	0.23	&	10.64	&	0.53	&	2	\\
	&	328.0	&	0.264	&	5.70	&	0.29	&	4.44	&	0.22	&	10.14	&	0.51	&	2	\\
	&	344.0	&	0.277	&	5.54	&	0.28	&	4.36	&	0.22	&	9.89	&	0.49	&	2	\\
	&	360.0	&	0.741	&	5.45	&	0.27	&	4.21	&	0.21	&	9.66	&	0.48	&	2	\\
	&	392.0	&	0.344	&	5.15	&	0.26	&	4.04	&	0.20	&	9.19	&	0.46	&	2	\\
	&	408.0	&	0.720	&	5.07	&	0.25	&	3.96	&	0.20	&	9.03	&	0.45	&	2	\\
	&	424.0	&	0.348	&	4.83	&	0.24	&	3.80	&	0.19	&	8.63	&	0.43	&	2	\\
	&	1477	&	0.102	&	1.96	&	0.04	&	1.52	&	0.03	&	3.49	&	0.07	&	3	\\

3C452	&	116.9	&	1.46	&	44.27	&	4.43	&	41.65	&	4.17	&	85.92	&	8.59	&	1	\\
	&	121.8	&	1.41	&	43.39	&	4.34	&	40.96	&	4.10	&	84.85	&	8.49	&	1	\\
	&	124.7	&	1.37	&	42.56	&	4.26	&	40.67	&	4.07	&	83.23	&	8.32	&	1	\\
	&	128.6	&	1.28	&	41.87	&	4.19	&	39.90	&	3.99	&	81.77	&	8.18	&	1	\\
	&	132.5	&	1.21	&	41.28	&	4.13	&	39.55	&	3.96	&	80.83	&	8.08	&	1	\\
	&	136.4	&	1.23	&	40.06	&	4.01	&	38.88	&	3.89	&	78.94	&	7.89	&	1	\\
	&	140.3	&	1.05	&	39.71	&	3.97	&	38.49	&	3.85	&	78.20	&	7.82	&	1	\\
	&	152.0	&	1.08	&	38.07	&	3.81	&	37.10	&	3.71	&	75.17	&	7.52	&	1	\\
	&	155.9	&	1.00	&	37.64	&	3.76	&	36.66	&	3.67	&	74.30	&	7.43	&	1	\\
	&	159.9	&	0.947	&	37.15	&	3.72	&	36.30	&	3.63	&	73.45	&	7.35	&	1	\\
	&	280.0	&	0.696	&	21.37	&	1.07	&	20.18	&	1.01	&	41.55	&	2.08	&	2	\\
	&	296.0	&	0.458	&	20.79	&	1.04	&	19.69	&	0.98	&	40.48	&	2.02	&	2	\\
	&	312.0	&	0.394	&	19.90	&	1.00	&	18.86	&	0.94	&	38.76	&	1.94	&	2	\\
	&	328.0	&	0.412	&	19.04	&	0.95	&	18.02	&	0.90	&	37.06	&	1.85	&	2	\\
	&	344.0	&	0.392	&	18.28	&	0.91	&	17.29	&	0.86	&	35.57	&	1.78	&	2	\\
	&	360.0	&	0.682	&	17.48	&	0.87	&	16.45	&	0.82	&	33.93	&	1.70	&	2	\\
	&	392.0	&	0.438	&	16.45	&	0.82	&	15.52	&	0.78	&	31.97	&	1.60	&	2	\\
	&	408.0	&	0.763	&	16.12	&	0.81	&	15.24	&	0.76	&	31.36	&	1.57	&	2	\\
	&	424.0	&	0.478	&	15.48	&	0.77	&	14.59	&	0.73	&	30.07	&	1.50	&	2	\\
	&	456.0	&	0.857	&	14.53	&	0.73	&	13.66	&	0.68	&	28.19	&	1.41	&	2	\\
	&	606.3	&	0.825	&	11.55	&	1.16	&	11.14	&	1.11	&	22.69	&	2.27	&	2	\\
	&	1413	&	0.209	&	5.32	&	0.11	&	5.21	&	0.10	&	10.54	&	0.21	&	4	\\
																																		
\hline

\end{tabular}

\vskip 5pt
\begin{minipage}{13.5cm}
`Frequency' refers to the frequency of the map. `Off-source RMS' refers to RMS noise measured over a large region well away from the source. `Integrated flux' values are listed at each frequency for the two lobes and for the whole source, where `Lobe 1' refers to the northern and eastern and `Lobe 2' to the southern and western lobes of 3C223 and 3C452 respectively. The associated uncertainties are the combined RMS and flux calibration errors. `Reference' lists the origin of the images used to measure each value as follows: (1) \citet{harwood16}; (2) This paper; (3) \citet{leahy91a}; (3) atlas of DRAGNs  (Leahy, Bridle \& Strom, \url{http://www.jb.man.ac.uk/atlas/}).
\end{minipage}
\end{table*}

\section{Data Reduction and Spectral Analysis}
\label{method}

\subsection{Target selection and data reduction}
\label{datareduction}

To investigate the dynamics of powerful radio galaxies at low frequencies we use the sample presented by H16 in the first paper of this series of two nearby FR IIs: 3C223 and 3C452 (Table 1). These sources were observed as part of the LOFAR surveys key science Project (KSP) and comprise of 10 hour observations of each source at LBA (10 - 90 MHz) and HBA (110 - 250 MHz) frequencies using the Dutch stations, giving a maximum baseline of $\sim$100 km. As within this paper we are interested in the spectrum and dynamics on well resolved scales, we limit ourselves to the higher frequency HBA observations which provide a spatial resolution of $\sim$7 arcsec. As we use the same data as H16 we do not repeat details of the reduction process here, but note that due to ionospheric effects and heavy flagging on the longest baselines in certain bands, the 3C223 data were reimaged at 13 arcsec resolution in order to obtain good image fidelity on small scales whilst maintaining the maximum possible frequency coverage.

To tightly constrain any curvature present in the spectrum of the sources at low frequencies, VLA P-band (224 - 480 MHz) observations were undertaken. To recover the most diffuse emission and obtain a resolution comparable to that of LOFAR, observations were made in A and B array configurations for 3C223 with an additional pointing in C-configuration for 3C452, using the full  256 MHz bandwidth centred at a frequency of 352 MHz. The data were reduced using \textsc{casa} in the standard manner and following the guidelines set out in the online tutorial for continuum P-band data\footnote{\burl{https://safe.nrao.edu/wiki/bin/view/Main/HuibIntemaPbandCasaGuide}}. We note that as the P-band receivers were recently upgraded the data were inspected for swapped polarizations, although none were found.

In order to apply the small scale spectral analysis described below in Section \ref{spectralanalysis}, multiple images were required across the observed frequency range. The full bandwidth was therefore split on a spectral window basis (16 MHz per SPW) giving a total of 16 images. To reduce computational overhead during imaging, each SPW was averaged down in frequency to 16 channels of 1 MHz bandwidth. No averaging in time was undertaken.

\begin{figure}
\hspace{-2mm}
\includegraphics[angle=0,width=8.5cm]{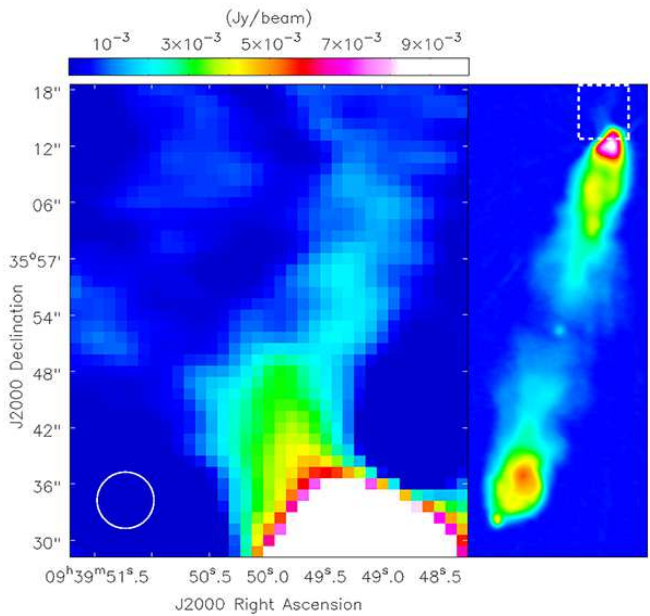}
\caption{The northern extension observed in 3C223 at 368 MHz. The dashed region of the full source (right) indicated the zoomed region (left) of the combined bandwidth image. The colour scale of the zoomed region set to highlight the diffuse, low surface brightness emission. The restoring beam of 6 arcsec is indicated in the bottom left corner of the image.}
\label{northernextension}
\end{figure}

To allow combination of the different array configurations, phase only self-calibration was undertaken on the the A-configuration data before being \textsc{clean}ed to convergence. These images were then used as a model in order to cross-calibrate the B-configuration data and, for 3C452, the combined A- and B-configuration images used to calibrate the C-configuration data. The multiscale clean algorithm (\textsc{msclean}, \citealp{cornwell08, rau11}) was used to produce all images at scales of 0, 5, 15 and 45 times the cell size, where the cell size was set to one-fifth of the beam size. To perform the required spectral analysis, the P-band imaging parameters must be matched to those of the LOFAR maps. We therefore use the same image size and resolve the beam of each map to the lowest LOFAR frequency as defined by H16. In addition, radio maps were also made using the full bandwidth at a central frequency of 368 MHz. As this required imaging over a large fractional bandwidth, we used the \textsc{casa} multifrequency synthesis \textsc{clean} parameter \emph{nterms} = 2 \citep{rau11} which determines the spectral index of the source over the observed bandwidth in order to scale the observed flux to the central frequency and produce a more realistic image of the target. The resulting images are shown in Figure \ref{pbandimages}.

To constrain curvature at higher frequencies, archival VLA images from the atlas of DRAGNs (Leahy, Bridle \& Strom\footnote{\url{http://www.jb.man.ac.uk/atlas/}}; \citealp{leahy91a}) were also obtained. As these images were already of a high quality, only regridding to J2000 coordinates and matching to the LOFAR and VLA P-band imaging parameters was required. For 3C452, the 610 MHz GMRT image produced by H16 was also included in our analysis. While other archival data were also available, they are unable to match the images described above in terms of either resolution, \emph{uv} coverage, or sensitivity. A summary of all of the data used (both new and archival) is given in Table \ref{imagedetails}.

While all of the images are matched in terms of sky coordinates, small frequency dependent shifts during calibration and imaging mean that an accurate alignment in image space is crucial to performing the small scale spectral analysis used in this paper. We therefore align the images using the Gaussian fitting method detailed by H13 and H15 which has previously proved successful for this form of analysis. As the radio core is not visible at all frequencies, Gaussians were instead fitted to point sources close to the target using \textsc{casaviewer}. A suitable reference pixel was then chosen and each image aligned using the \textsc{aips}\footnote{\url{http://www.aips.nrao.edu/index.shtml}} OGEOM task. To check the accuracy of the alignment, Gaussians were again fitted to the resulting images where we found a difference between the maps of only around 0.01 pixels. We can therefore be confident that the effects of misalignment are likely to be negligible in our analysis.

\subsection{Spectral analysis}
\label{spectralanalysis}

In order to achieve the aims of this paper, we have used the Broad Radio Astronomy ToolS\footnote{\url{http://www.askanastronomer.co.uk/brats}} software package (\textsc{brats}) which provides a suite of tools designed for the fitting of spectral ageing models and spectral indices on a range of spatial scales, as well as various statistical and mapping tools. Details of the methods underlying the software have been discussed extensively by H13 and H15 and, as our analysis follows the standard procedures laid out in the \textsc{brats} cookbook\footnote{\url{http://www.askanastronomer.co.uk/brats/downloads/bratscookbook.pdf}}, we do not repeat such a discussion here but instead highlight the specific functions and parameters used for the analysis of our data. 

\begin{figure*}
\centering
\includegraphics[angle=0,width=17.6cm]{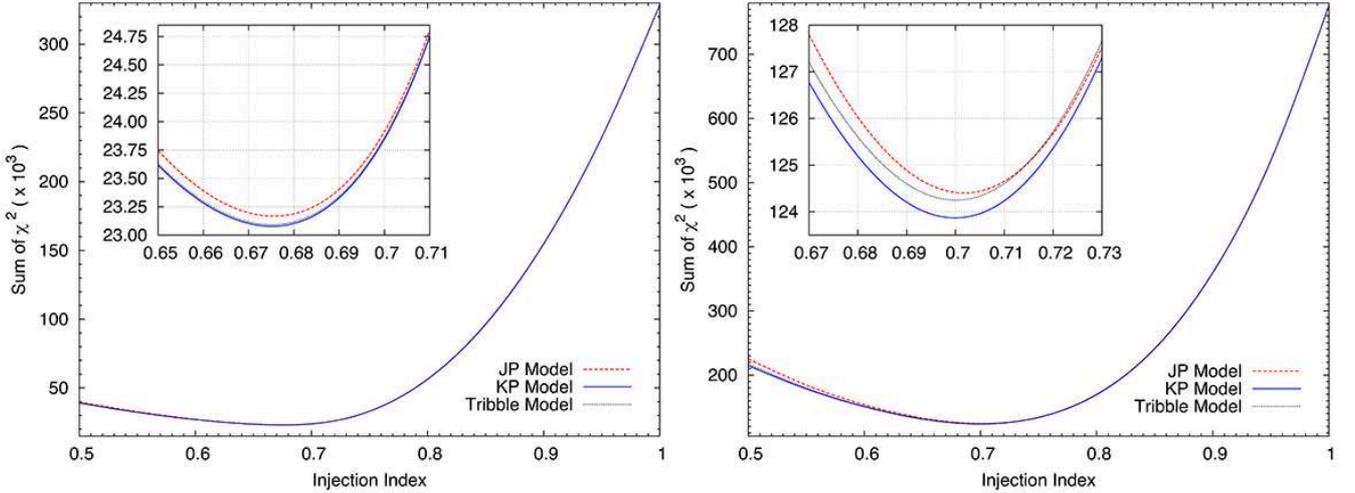}
\caption{$\chi^{2}$ values for 3C223 (left) and 3C452 (right) for the JP (red), KP (blue) and Tribble (black) models of spectral ageing as a function of injection index. Plots zoomed to the minimum values are shown inset for each panel. The minimum occurs at 0.68 for 3C223 and 0.70 for 3C452 to 2 significant figures for all models. Note that for 3C223 the Tribble model closely follows KP model. Fitting was performed with an initial step of 0.05 between 0.5 and 1.0 and a secondary step size of 0.01 around the minimum values. The data are fitted with natural cubic splines. As all points lie on the fitted spline they are excluded for clarity.}
\label{inject}
\end{figure*}

In order to ensure only the target source was included in our analysis, and to reduce computational overheads, a region that loosely encompasses the source was initially defined for both 3C452 and 3C223 using \textsc{ds9}\footnote{\url{http://ds9.so.edu}}. The core of each target, which is not expected to be described by models of spectral ageing, was excluded in both cases so as that our analysis is not biased by this region. To determine the off-source thermal noise a background region was also defined for each field in a blank area of sky well away from the target. The data were then loaded in to \textsc{brats} and an initial source detection performed based on a 5$\sigma$ cutoff for each image. To account for the increased RMS noise associated with uncertainty in the modelling of extended emission during imaging, we adopt an on-source noise multiplier of 3 as described by H13 and H15 for determining statistical values and for the final region selection process. For the uncertainty introduced due to flux calibration errors we have assumed at LOFAR frequencies the value of 10 per cent presented by H16, for GMRT observations a 5 per cent uncertainty \citep{scaife12}, and for VLA observations the standard values of 5 and 2 per cent at P- and L-band respectively \citep{scaife12, perley13}. To reduce the effects of the superposition of spectra (H13; H15), the \textsc{brats} `\emph{setregions}' command was set to define regions on a pixel by pixel basis, resulting in a total of 9272 regions for 3C452 and 2620 regions for 3C223.

To determine the age of the observed emission, three models of spectral ageing were fitted to the spectrum of both 3C452 and 3C223. The first two are the standard models proposed by \citet{kardashev62} and \citet{pacholczyk70} (KP model) and by \citet{jaffe73} (JP model) which assume a single initial injection of particles described by a power law (Equations \ref{initialdist} \& \ref{alphainject}) which are then subject to radiative losses via the synchrotron process and inverse-Compton scattering on the cosmic microwave background (CMB). These models differ only in that the KP model assumes a fixed pitch angle of the electrons with respect to the magnetic field whereas the JP model assumes a variation in this angle, averaged over it's radiative lifetime. As a consequence, the KP model displays much less curvature for an equivalent age than the exponential cutoff seen in the JP model, due to there always being a supply of high energy electrons maintaining a small pitch angle to the magnetic field, hence lower losses at higher frequencies.

While both the JP and KP model assume the electrons are subject to a uniform magnetic field over their lifetimes, the third model tested attempts to account for a potentially more realistic scenario in which the field is spatially non-uniform. First proposed by \citet{tribble93}, the requirement for ageing models to be fast fitting (on the order of a few seconds) so that the parameter space can be fully explored and the need to be well constrained in frequency space to make a reasonable comparison to the standard models has historically limited its application to observations; however, it has recently been shown that in the free-streaming (i.e. weak field, high diffusion) case the observed spectrum is simply the JP model integrated over the Maxwell-Boltzmann distribution and so can be comfortably handled by current computational capabilities (\citealp{hardcastle13a}; H13). Using the new generation of broad bandwidth interferometer, this model has proved successful in combining the physical plausibility of the time averaged pitch angle provided by the JP model with the often better goodness-of-fit provided by the KP model (e.g. H15), but has yet to be investigated at low frequencies.

In the fitting of these models, we have used the magnetic field strengths determined by H16 from synchrotron/inverse-Compton fitting of 0.45 nT and 0.34 nT for 3C452 and 3C223 respectively and assume standard values of $\gamma_{min} = 10$ and $\gamma_{max} = 1 \times 10^{6}$ for the minimum and maximum Lorentz factors respectively (\citealp{carilli91, hardcastle98, godfrey09}; H16). In order to determine the injection index we use the \textsc{brats} `\emph{mininject}' command. By setting the injection index as a free parameter, this function uses $\chi^{2}$ minimisation to determine which injection index best describes the source and allows any curvature found within the spectrum due to ageing to be accounted for (see H13 for a detailed description). We performed a grid search between $\alpha_{inj} = 0.5$ and a reasonable upper limit of $\alpha_{inj} = 1.0$ using an initial step size of 0.05.  A second fitting was then performed for each source around the minimum value using a decreased step size of 0.01. Fitting of the three models described above was then performed and the relevant images, data, and statistical values exported for analysis. Note that the analysis considers only the current episode of activity. The potential impact of multiple episodes on the intrinsic age of radio galaxies is discussed further in Section \ref{dynamicalmodel}.

\section{Results}
\label{results}

\subsection{Data quality}
\label{dataquality}

Of the 16 spectral windows that make up the new VLA P-band observations, 10 SPWs for 3C452 and 9 SPWs for 3C223 were found to have high quality data in all array configurations. The spectral windows between 224 and 272 MHz (SPWs 0 to 2), 376 MHz (SPW 9) and 472 MHz (SPW 15) are purposely positioned at frequency ranges heavily affected by known RFI sources and so were of extremely poor quality for both targets in all array configurations and were therefore excluded from our analysis. At 440 MHz (SPW 13, both sources) and 456 MHz (SPW 14, 3C223 only), heavy flagging was required in at least one array configuration and so these data were also excluded from consideration.

\subsection{Morphology and size}
\label{morphology}

From the resulting combined P-band images (Fig. \ref{pbandimages}) we see that, as expected, the large scale morphology matches that of previous studies (e.g. \citealp{nandi10, orru10}) and of H16 at LOFAR frequencies. Measuring the largest angular extent of the sources from the 3 sigma contour at the tips of each lobe (Table \ref{jvlatargets}) we find an increase in size compared to those of the 3CRR catalogue \citep{laing83}. 3C452 increases only marginally from 428 to 438 kpc with a greater difference observed in 3C223 which increases from 740 to 796 kpc, placing it comfortably within the giant radio galaxy classification range.

While the large scale structure is generally in good agreement with previous findings one notable exception is a faint, approximately 30 arcsec (72 kpc) northern extension to 3C223 (Fig. \ref{northernextension}). The faint nature of this emission means that it is not visible in archival images at GHz frequencies but hints of this structure were originally evident in the LOFAR image of H16. As it has a surface brightness around the noise level of the LOFAR image and is not observed at higher frequencies, it could not be confidently confirmed; however, this emission is clearly visible in the VLA P-band images at a location coincident with the tenuous detection in the LOFAR maps and so is unlikely to be an imaging artefact. Taking the 3 sigma noise local to northern lobe from the full bandwidth LOFAR HBA image at 368 MHz of H16 and correcting for the differing beam sizes, we can place upper limits on the spectral index of $\alpha < 1.5$ for emission closest to the tip of the lobe and $\alpha < 2.0$ at its furthest extent. While not particularly constraining, these upper limits suggest that the emission is not unphysically steep. No optical counterpart is found in the SDSS catalogue \citep{alam15} and so, given the morphology, it is reasonable to assume that it is associated with 3C223. The underlying cause of this extended structure is unclear but may be due to a low density region of the surrounding medium which has subsequently filled with plasma from the radio galaxy lobes. We note that as this structure is not visible at higher frequencies and is tenuous in the LOFAR image, it has not been included in the determination of the source size or in our ageing analysis.

\begin{figure*}
\centering
\includegraphics[angle=0,height=22.2cm]{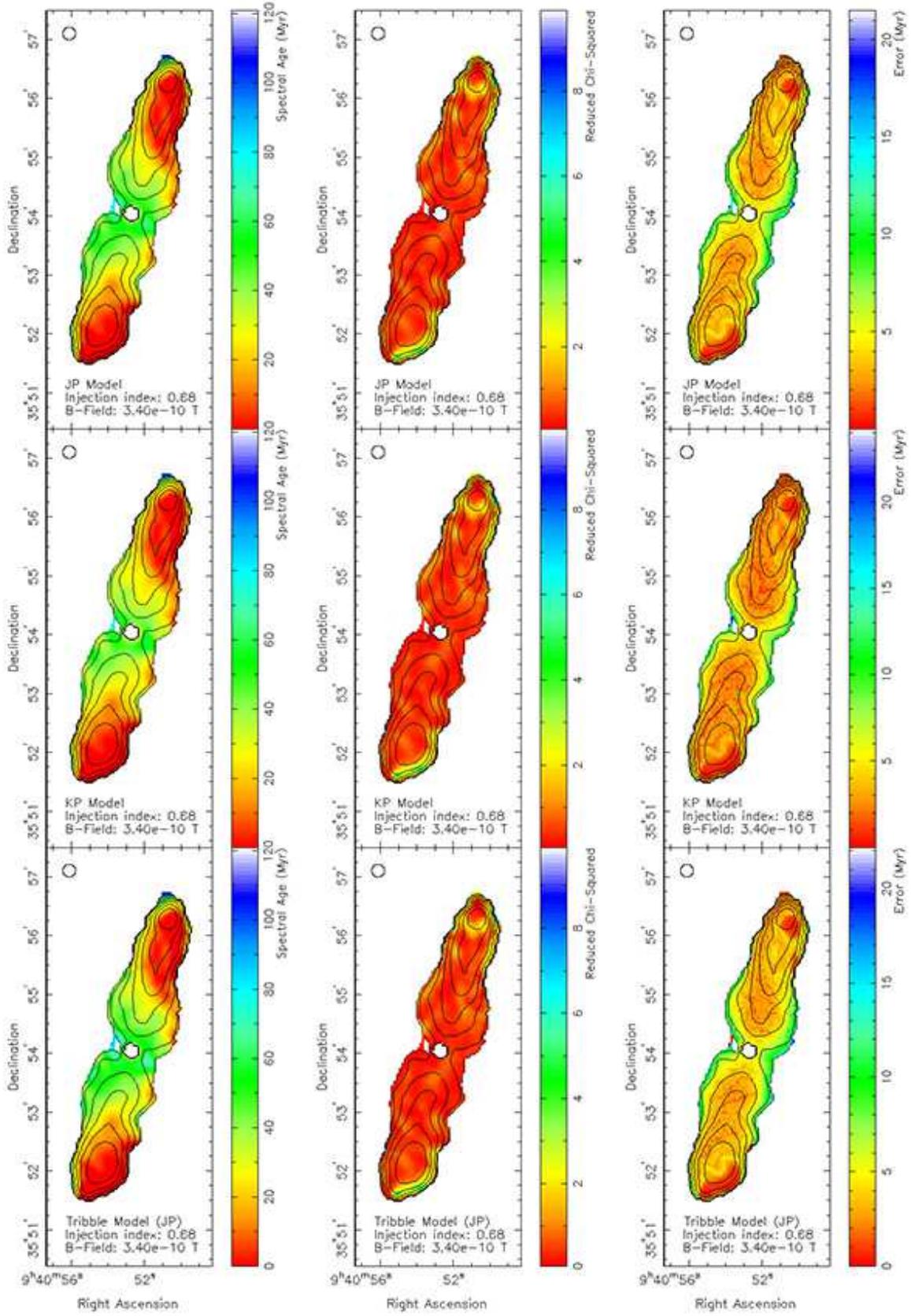}
\caption{Spectral ageing maps of 3C223 (left) with corresponding $\chi^{2}$ (middle) and error maps (right) for the JP (top), KP (middle) and Tribble (bottom) models of spectral ageing. All maps have been overlaid with 1.4 GHz flux contours. A magnetic field strength of 0.34 nT \citep{harwood16} was used for all fits along with an injection index of 0.68 as described in Section \ref{specageresults}. The maximum spectral ages of all models given in Table \ref{lobespeed}.}
\label{3C223_specage}
\end{figure*}

\subsection{Model fitting and parameters}
\label{specageresults}

The results of the injection index determination are shown in Figure \ref{inject}. Both sources have a similar minimum of $\alpha_{inj} = 0.70 \pm 0.01$ and $0.68 \pm 0.01$ for 3C452 and 3C223 respectively with all models converging to their respective values. 

While the recent study by H15 at GHz frequencies using the same methods applied here notes that such a determination is model dependent, so that some difference is expected, this convergence at LOFAR and P-band frequencies is not surprising. We see from Equation \ref{freqscale} that as energy losses scale as $\tau \propto 1/\nu^{2}$, low frequency observations are less affected by curvature due to spectral ageing for all but the oldest regions of plasma, hence all models should converge to a power law described by the injection index. It is therefore consistent with what one would expect for such low frequency observations; however, the observed difference between the injection index determined from the integrated spectrum by H16 ($0.85$ and $0.71$ for 3C452 and 3C223 respectively) and that derived here is much more unexpected.

\begin{figure*}
\centering
\includegraphics[angle=0,width=17.8cm]{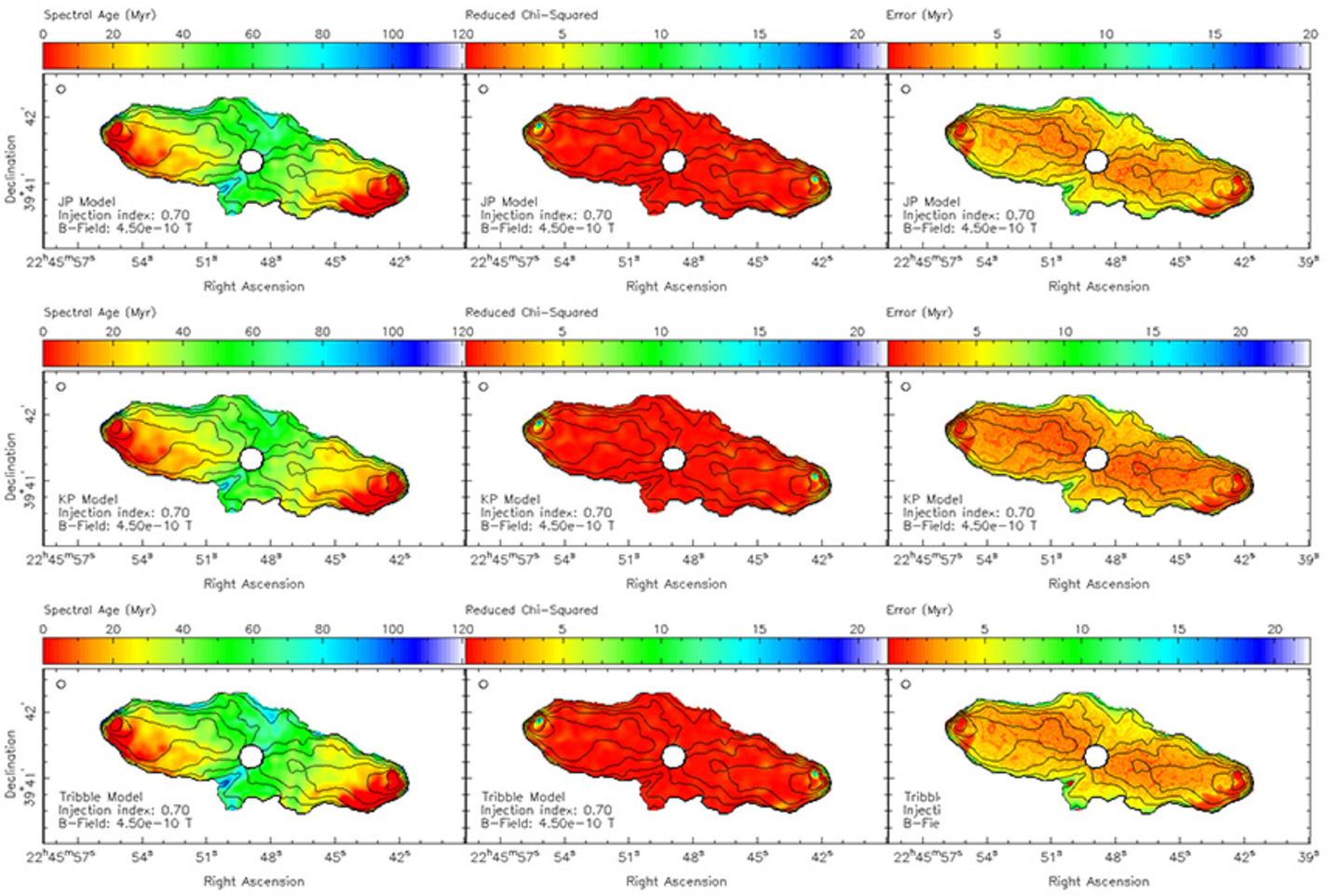}
\caption{Spectral ageing maps of 3C452 (left) with corresponding $\chi^{2}$ (middle) and error maps (right) for the JP (top), KP (middle) and Tribble (bottom) models of spectral ageing. All maps have been overlaid with 1.4 GHz flux contours. A magnetic field strength of 0.45 nT \citep{harwood16} was used for all fits along with an injection index of 0.70 as described in Section \ref{specageresults}. The maximum spectral ages of all models given in Table \ref{lobespeed}.}
\label{3C452_specage}
\end{figure*}

Comparing these values to those presented by H16, we see that the value determined from the integrated spectrum of 3C223 is only slightly steeper than the injection index presented here ($\Delta\alpha = 0.03$) and, given that the source is relatively old (Section \ref{specageresults}), is easily accounted for by curvature due to spectral ageing in the oldest regions of plasma. However, the large disparity between the value determined from the integrated spectrum of 3C452 and the injection index presented within this paper ($\Delta\alpha = 0.15$) cannot be explained simply via radiative losses in homogeneous models. A steeper injection index ($\alpha_{inj} = 0.78$) was also derived by \citet{nandi10} who use a similar method of setting $\alpha_{inj}$ as a free parameter but over spatially larger regions, resulting in a superposition of spectra which is known to have a significant impact on the observed spectral index (e.g. \citealp{stroe14, kapinska15}). As the method for determining the injection index used here should reduce such effects it is this value which we apply in the subsequent model fitting. We discuss in detail these results and the possible physical causes further in Section \ref{hotspots}.

\begin{figure*}
\centering
\includegraphics[angle=0,width=17.8cm]{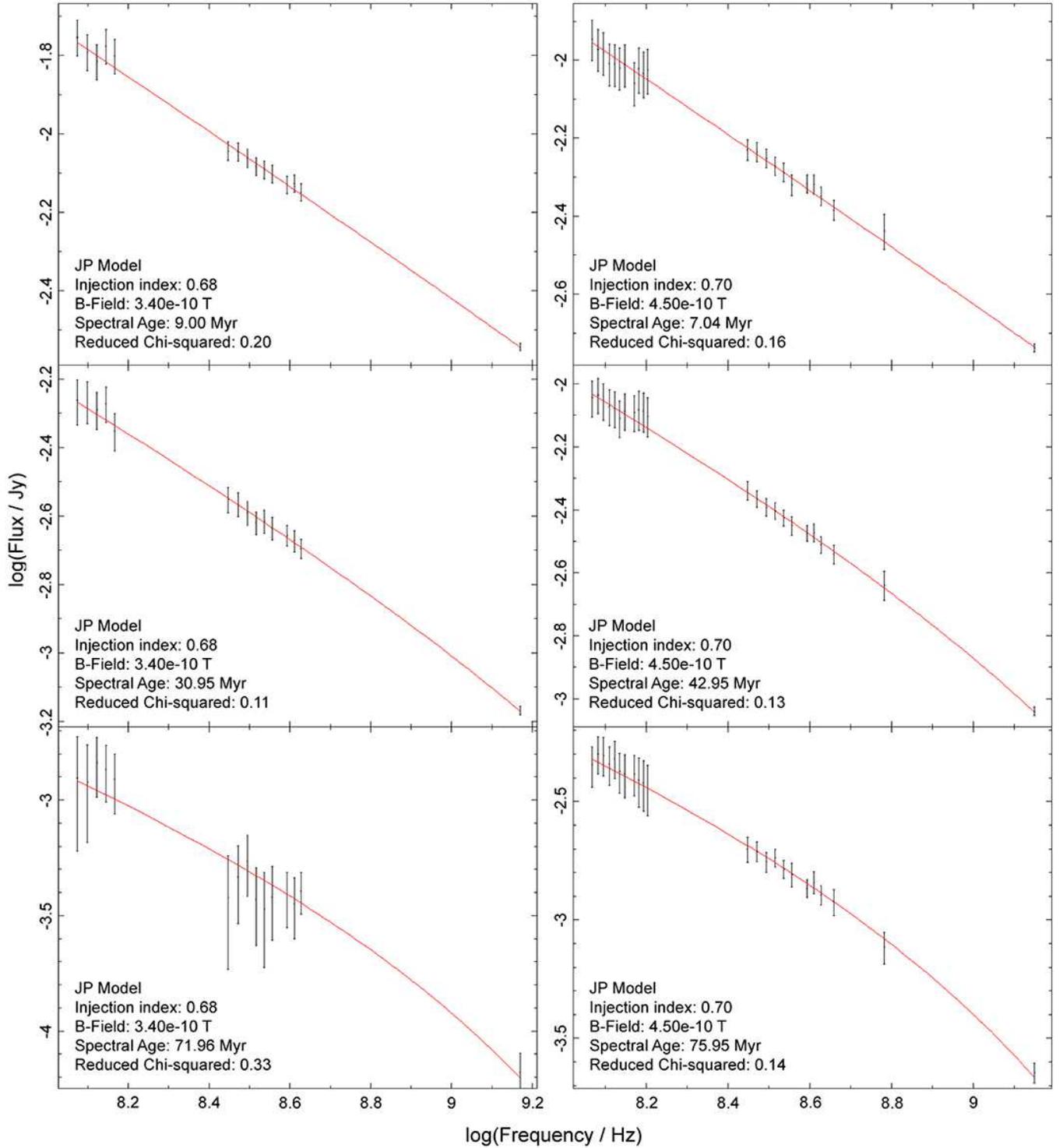}
\caption{Plots of flux against frequency for a selection of regions in 3C223 (left) and 3C452 (right) with the JP model of spectral ageing overlaid (red line). Three spectral age ranges are shown: low age (top), moderate age (middle), and high age (bottom). Model parameters, ages (in Myr), and statistics are shown in the bottom left corner of each panel.}
\label{exampleplots}
\clearpage
\end{figure*}

Figures \ref{3C452_specage} and \ref{3C223_specage} show the spectral ageing maps for 3C452 and 3C223 respectively, along with their corresponding $\chi^{2}$ and errors as a function of position. The general distribution of spectral ages for all three models is consistent with what one would expect for powerful FR IIs, with low age regions of plasma coincident with the hotspots and the oldest regions of plasma residing close to the AGN core. Plots of frequency against flux of representative regions for various ages are shown in Fig. \ref{exampleplots}.

From Table \ref{fittingresults}, which shows the statistical values for the model fitting, we see that all three models are able to provide a good description of the observed spectrum with none of the models being rejected at even the one sigma level. From Table \ref{lobespeed} we see that the spectral ages of the sources range from $77.05_{-8.74} ^{+9.22}$ (Tribble model) and $89.05_{-7.14}^{+8.56}$ (JP model) Myr for 3C452 and between $73.96_{-13.92}^{+17.29}$ (KP model) and $84.95_{-13.83}^{+15.02}$ (Tribble model) Myr for 3C223. The large error bars associated with the ages are unsurprising due to the limited number of data points at GHz frequencies where curvature due to ageing is most easily constrained, a point discussed further in Section \ref{specagediscussion}.

While the statistics of Table \ref{fittingresults} suggest that the models of spectral ageing tested provide a good description of the source as a whole, localised regions of high $\chi^{2}$ values that are spatially coincident for all three models are also observed. From the $\chi^{2}$ maps of Figures \ref{3C452_specage} and \ref{3C223_specage}, we see that these poorly fitted regions are mainly confined to two areas, the first of these being at the tip the lobes where the gradient in the flux is sharpest (Fig. \ref{pbandimages}). Such edge effects have previously been discussed by H13 and H15 and are likely a result of insufficient image fidelity in these regions and conclude that fitting here is unlikely to provide a robust measure of the age of the plasma and so are excluded from subsequent analysis.

The second of the poorly fitted regions are those located coincident with the hotspots, a behaviour which has also been observed in previous studies at GHz frequencies by H13 and H15. They therefore appear to be a fairly common feature when fitting to bright hotspots in FR IIs but previous studies have been unable to definitively determine the cause of this discrepancy due to these investigations being performed at a higher frequency range. The longer wavelengths used within this paper, where the effects of ageing should be negligible, allow us to probe the possible cause for these regions being poorly described by the models and we discuss this further in Section \ref{hotspots}, but we note here that it is likely to be a real feature of the spectrum, rather than due to any imaging effects.

Taking the ages provided by the Tribble model\footnote{We choose the Tribble model here as it has previously been shown (H13, H15) to provide the best balance between goodness-of-fit and physical plausibility, although the large errors associated with all of the models mean this selection makes only a marginal difference for these comparisons.} and comparing them to those of previous studies, the spectral age derived by \citet{nandi10} for 3C452 of only 27 Myr is significantly lower than the age derived within this paper. We discuss this discrepancy further in Section \ref{specagediscussion} but note here that it is likely due a combination of a steeper injection index, a magnetic field strength determined from equipartition, and the size of the regions considered in the original study.

The age of 3C223 provides much better agreement with those derived by \citet{orru10} who find an age of $72 \pm 4$ Myr, well within our margin of error. While they use similar methods to those of \citet{nandi10}, we see from Figure \ref{3C223_specage} that 3C223 has a far less cross-lobe age variations meaning that the superposition of spectra plays a much less significant role in determining the derived model parameters and age. Similarly, H16 find that the magnetic field strength is 75 per cent of the equipartition value and so remains relatively close to that assumed by \citet{orru10}. The ages presented within this paper for 3C223 are therefore in good agreement with the literature, although additional constraints at GHz frequencies are required if a more accurate age is to be determined. We discuss this further in Section \ref{specagediscussion}.

\begin{table*}
\centering
\caption{Model Fitting Results}
\label{fittingresults}
\begin{tabular}{ccccccccccc}
\hline
\hline

Source&Model&$\sum \chi^{2}$&Mean $\chi^{2}$&\multicolumn{5}{c}{Confidence Bins}&Rejected&Median\\
&&&&$<$ 68&68 - 90&90 - 95&95 - 99&$\geq$ 99&&Confidence\\

\hline
3C223	&	JP	&	23188	&	8.85	&	2285	&	102	&	41	&	80	&	112	&	No	&	$<$ 68	\\
	&	KP	&	23102	&	8.82	&	2281	&	106	&	39	&	80	&	114	&	No	&	$<$ 68	\\
	&	Tribble	&	23118	&	8.82	&	2285	&	102	&	41	&	78	&	114	&	No	&	$<$ 68	\\
3C452	&	JP	&	124415	&	13.42	&	8384	&	288	&	89	&	136	&	375	&	No	&	$<$ 68	\\
	&	KP	&	123870	&	13.36	&	8391	&	286	&	86	&	136	&	373	&	No	&	$<$ 68	\\
	&	Tribble	&	124250	&	13.40	&	8393	&	285	&	84	&	136	&	374	&	No	&	$<$ 68	\\
\hline
\end{tabular}

\vskip 5pt
\begin{minipage}{14.8cm}
`Model' refers to the spectral ageing model fitted to the target listed in the `Source' column.  $\sum \chi^{2}$ lists the sum of $\chi^{2}$ overall all regions with an equivalent mean value shown in the `Mean $\chi^{2}$' column. `Confidence Bins' lists the number of regions for which their $\chi^{2}$ values falls with the stated confidence range. `Rejected' lists whether the goodness-of-fit to the source as a whole can be rejected and `Median Confidence' the confidence level at which the model can be rejected.
\end{minipage}
\end{table*}

 \subsection{Lobe advance speeds}
\label{advancespeeds}

In order to determine the speed at which the lobes of our sample are advancing through the external medium we employ the standard method of \citet{alexander87}. Assuming the source is orthogonal to the line of site and that hotspots are the primary sites for particle acceleration, a characteristic advance speed $v_{lobe}$ is given simply by $v_{lobe} = d_{hs} / t_{s}$ where $d_{hs}$ is the distance between the core and the hotspot, and $t_{s}$ is the age of the source. Within this paper we define hotspots to be the bright, flat spectrum emission located near the tip of the lobes where particle acceleration is most likely to be taking place, although acknowledge this does not necessarily fit the strict morphological definition laid out by e.g. \citet{leahy97} who define hotspots as a single distinct feature with its largest dimension less than 10 per cent of the lobe's major axis, a peak brightness 10 times the RMS noise, and separated from any nearby  peaks by a minimum falling two-thirds of less of the fainter peak.

Applying this to our sample, we find that 3C223 is the more rapidly expanding with the northern lobe advancing between $1.33_{-0.05}^{+0.06}$ (Tribble) and $1.53_{-0.06}^{+0.08}$ (KP) $\times 10^{-2}\,c$ and the southern lobe between $1.37_{-0.05}^{+0.06}$ (Tribble) and $1.58_{-0.06}^{+0.08}$ (KP) $\times 10^{-2}\,c$. 3C452 is advancing at roughly half the speed of 3C223, with the eastern lobe between $0.71_{-0.03}^{+0.03}$ (JP) and $0.82_{-0.04}^{+0.04}$ (Tribble) $\times 10^{-2}\,c$ and the western lobe between $0.73_{-0.03}^{+0.03}$ (JP) and $0.84_{-0.04}^{+0.04}$ (Tribble) $\times 10^{-2}\,c$. A summary of the derived characteristic advance speeds for each source and lobe is given in Table \ref{lobespeed}. 

Previous studies of similar, well established FR IIs \citep{myers85, alexander87, liu92, odea09} find a range of advance speeds from a few to tens of percent the speed of light. 3C452 and 3C223 therefore fall at the lower end of this scale; however, a direct comparison may not be appropriate in this case. Previous studies make the assumption that, at low frequencies, the integrated spectrum is a superposition of power law spectra with negligible curvature due to ageing and therefore provides a robust measure of the injection index. In light of the results of Section \ref{specageresults} where we find that the integrated spectrum has the potential to be significantly steeper than the resolved case, and of \cite{croston05} and H16 who show that the magnetic field strength of these sources is generally less than equipartition (also see examples by \citealp{brunetti02, isobe05, migliori07}), it is likely that the spectral age of the sources in these previous studies are underestimated leading to an increased advance speed when compared to well resolved studies using modern techniques.

Recent investigations of two FR IIs (3C28 and 3C438) by H15 where the injection indices were determined on resolved scales (although equipartition was still assumed) find advance speeds of around 2 per cent the speed of light, more in line with those of 3C452 and 3C223, but larger samples using resolved techniques are required to conclusively determine whether the characteristic advance speed of FR IIs is lower than previously thought.

\section{Discussion}
\label{discussion}

The results presented in Section \ref{results} provide the first opportunity to accurately determine in detail the spectrum of FR IIs at low frequencies on small spatial scales and investigate the particle acceleration processes which drive their emission, vital if we are to determine the life-cycle of radio galaxies and their impact on the external environment. In the following sections we discuss the reliability of our results and give comparison to previous studies and how these findings effect our understanding of the observed spectrum of FR IIs at low frequencies and the dynamics of these powerful radio sources.

\subsection{Initial electron energy distribution and particle acceleration}
\label{hotspots}

One of the key recent findings with respect to the parameters of spectral ageing models is that, when tightly constrained over a wide frequency range on small spatial scales, the observed injection index is significantly steeper than the values of around 0.5 to 0.6 which has been traditionally assumed.

\begin{figure*}
\centering
\includegraphics[angle=0,width=18.0cm]{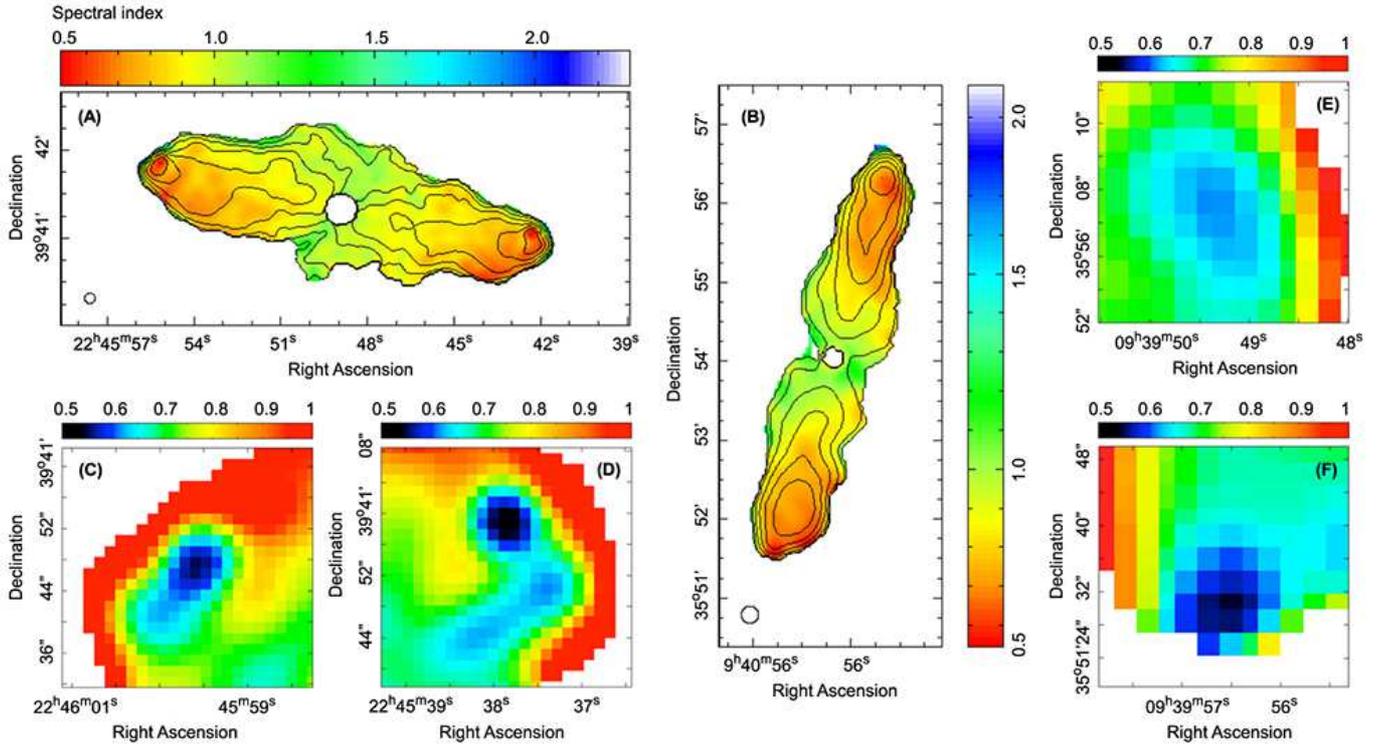}
\caption{Spectral index maps of 3C452 (A) and 3C223 (B) with 1.4 GHz flux contours overlaid and a standard min--max colour scale. The restoring beam of 6 and 13 arcsec for 3C452 and 3C223 respectively are indicated in the bottom left corner of the images. Panels (C) through (F) are zoomed to regions surrounding the hotspots for each source with the colour scale set to highlight the differences in these spectrally flatter regions of plasma. Note the jump between $\alpha \approx 0.5$ and $\alpha \approx 0.7$ is not due to curvature, as discussed in Section \ref{hotspots}.}
\label{specindex}
\end{figure*}

With the exception of 3C28 which is thought to be an inactive, remnant radio galaxy, the injection index values are consistent with the previous studies on small spatial scales by H13 and H15 who find values of around 0.7 to 0.8. Interestingly, this is only slightly lower than the average integrated spectral index between 178 and 750 MHz of $\alpha \approx 0.8$ of the 3CRR sample by \citet{laing83}, suggesting that the model fitting methods used both here and by H13 and H15 are successfully accounting for curvature due to ageing and reliably tracing the low frequency spectrum as expected. However, from Figure \ref{specindex}, which shows the spectral index of 3C452 and 3C223 as a function of position (weighted least squares fit to all data points), it is unclear whether this is truly tracing the initial electron energy distribution.

As discussed in Section \ref{specageintro}, the injection index is related to the power law index which describes the initial electron energy distribution by Equation \ref{alphainject} which is then subject to radiative losses, in theory leading to a smooth gradient of spectra between the initial acceleration at the hotspots and the oldest regions of plasma close to the core. From panels C through F of Fig. \ref{specindex} which show the regions of plasma close to the hotspot we clearly see, in contradiction to this assumption, a jump in the spectral index from between $0.5$ and $0.6$ (the historically assumed values) in the hotspots and $\gtrsim\,$$0.7$ (those derived from model fitting and the integrated spectrum) just outside on angular scales comparable to the resolution of the radio images. For 3C452, these scales can plausibly be increased due to the extended flat spectrum structure connected to the hotspot being a back flow of material which, while not explaining the cause of the steepening, would allow additional time for such a process to occur. However, this cannot be the case for 3C223 where such an extension is not present hence this  rapid spectral steepening must either occur on physical scales less than approximately 10 and 30 kpc for 3C452 and 3C223 respectively or from additional processes modifying the observed spectrum of the hotspots relative to that of the lobes.

\begin{table*}
\centering
\caption{Lobe Advance Speeds}
\label{lobespeed}
\begin{tabular}{lcccccccccc}
\hline
\hline
Source&Model&Lobe&Max Age&+&-&Angular Size&Distance&Speed&+&-\\
&&&(Myrs)&&&('')&(kpc)&($10^{-2}$ $v/c$)&&\\
\hline
3C223	&	JP	&	North	&	77.95	&	13.42	&	11.73	&	165.64	&	346.45	&	1.450	&	0.056	&	0.049	\\
	&	KP	&	North	&	73.96	&	17.29	&	13.92	&	163.20	&	346.45	&	1.528	&	0.076	&	0.061	\\
	&	Tribble	&	North	&	84.95	&	15.02	&	13.83	&	163.20	&	346.45	&	1.330	&	0.058	&	0.053	\\
	&	JP	&	South	&	77.95	&	13.42	&	11.73	&	163.20	&	357.12	&	1.494	&	0.056	&	0.049	\\
	&	KP	&	South	&	73.96	&	17.29	&	13.92	&	163.20	&	357.12	&	1.575	&	0.076	&	0.061	\\
	&	Tribble	&	South	&	84.95	&	15.02	&	13.83	&	163.20	&	357.12	&	1.371	&	0.058	&	0.053	\\
3C452	&	JP	&	East	&	89.05	&	8.56	&	7.14	&	139.32	&	193.74	&	0.710	&	0.031	&	0.026	\\
	&	KP	&	East	&	85.99	&	9.70	&	7.90	&	139.32	&	193.74	&	0.735	&	0.037	&	0.030	\\
	&	Tribble	&	East	&	77.05	&	9.22	&	8.74	&	139.32	&	193.74	&	0.820	&	0.039	&	0.037	\\
	&	JP	&	West	&	89.05	&	8.56	&	7.14	&	146.88	&	197.86	&	0.725	&	0.031	&	0.026	\\
	&	KP	&	West	&	85.99	&	9.70	&	7.90	&	146.88	&	197.86	&	0.750	&	0.037	&	0.030	\\
	&	Tribble	&	West	&	77.05	&	9.22	&	8.74	&	146.88	&	197.86	&	0.838	&	0.039	&	0.037	\\
\hline
\end{tabular}

\vskip 5pt
\begin{minipage}{15.1cm}
`Model' is the spectral ageing model fitted to the target listed in the `Source' column. `Max Age' is the maximum age of the corresponding `Lobe' in Myrs. `Angular size' gives the separation between the core and the hotspot in arcsec with the corresponding physical size given in the `Distance' column in kpc. `Speed' lists the derived lobe advance speed as a fraction of the speed of light as detailed in Section \ref{advancespeeds}. 
\end{minipage}
\end{table*}

As previously mentioned in Section \ref{specageresults}, the effects of spectral ageing are significantly reduced when observed well below the break frequency, $\nu_{t}$, given in GHz by \begin{equation}\label{specbreak}\nu_{t} = \frac{(9/4)\,c_7\,B}{(B^{2} + B_{CMB}^{2})^{2}\,t^{2}}\end{equation} where $t$ is the age of the source in Myr, $B$ is the magnetic field strength in nT, $B_{CMB} = 0.318(1+z)^{2}$ nT is the equivalent magnetic field strength for the CMB and $c_7 = 1.12 \times 10^3$ nT$^{3}$ Myr$^{2}$ GHz is a constant defined by \citet{pacholczyk70}. For a maximum spectral age of 80 Myr, the break frequency in any region of plasma within our observations must be above 1.53 and 1.65 GHz for 3C452 and 3C223 respectively, suggesting that spectral curvature should play a negligible role at LOFAR and P-band wavelengths for the sources in our sample, particularly when close to the acceleration regions. This appears to be borne out in our spectral fitting where both sides of the discontinuity are well fitted by a power law (Fig. \ref{specindex}) with no significant signs of curvature (zero age emission of Fig. \ref{3C452_specage} and \ref{3C223_specage}) and can therefore be ruled out as the cause of the spectral steepening.

One factor which is currently unaccounted for in spectral ageing models is adiabatic expansion. This process is well known to occur in hotspots, reducing the value of $\gamma_{min}$ from possibly as high as 1000 \citep{carilli91, hardcastle98} to $\sim$10 \citep{godfrey09} for the lobes; however, while adiabatic expansion can cause the break in a curved spectrum to move to lower frequencies, the lack of curvature in these regions means that adiabatic expansion alone will only reduce the intensity of the emission, not increase the steepness of its spectrum. We can therefore confidently rule out adiabatic expansion in homogeneous models alone as the cause of the observed disparity.

Another possible explanation is that the observed spectrum results from the combination of adiabatic expansion, strong magnetic field strengths, and the superposition of a range of high energy cutoffs. The observed emission of the recently accelerated plasma at any given frequency is given by\footnote{Note that the Tribble model is additionally integrated over the Maxwell-Boltzmann distribution. See \citet{hardcastle13a} and \citet{harwood13} for further details.}  \begin{equation}\label{flux}S(\nu) = \int^{\pi}_{0} \int^{E_{max}}_{E_{min}}\, J(\nu) \,N(E)\, dE\, d\theta \end{equation} where $N(E)$ is the initial electron energy distribution given by Equation \ref{initialdist} and $J(\nu)$ is the emissivity of a single electron give by \begin{equation}\label{emiss}J(\nu) = \frac{\sqrt{3} e^{3}} {8 \pi \epsilon_{0} c m_{e}}\, F(x) B \sin^2\theta \end{equation} where $B$ is the magnetic field strength, $\theta$ is the pitch angle of the electron to the magnetic field and $F(x)$ is a constant defined by \citet{pacholczyk70} where $x=\nu/\nu_{c}$ is a function of the critical frequency of the electron (see H13 for further details and \citeauthor{longair11}, \citeyear{longair11}, for a full derivation).

While it is generally assumed that we always observe a single high energy cutoff ($E_{max}$) for the plasma injected into the lobes this does not have to be the case. In principle, if the jet plasma is initially accelerated to the observed power law of $\alpha_{HS} \approx 0.5$ in the hotspot (as one would expect from e.g. first order Fermi acceleration, \citealp{blandford78}) but a significant variance in the high energy cutoffs are present, this may result in a steeper than expected spectrum once injected into the lobes. At the point of acceleration, all of these cutoffs are likely to be well above the observed frequency range but, as this region of frequency space is no longer a power law, they would be driven to lower energies by the rapid adiabatic expansion and strong magnetic field strengths known to exist in hotspots (e.g. \citealp{hardcastle98, hardcastle02b, araudo15}). As the plasma enters the lobe where the magnetic field strength and adiabatic expansion is relatively weak, these high energy cutoffs (prior to the effects of ageing) become fixed and the observed superposition of these spectra within any given beam may explain the observed rapid steepening of the power law spectrum. As a result, the $E_{max}$ term of Equation \ref{flux} would instead be the sum of a distribution of energy cutoffs, something which is currently not described by models of spectral ageing or radio galaxy dynamics.

However, while plausible, observations of hotspots at higher energies suggest such a superposition of cutoffs is unlikely to be the case for all sources due to the observed optical and infrared emission present in the hotspots of at least some FR IIs (e.g. \citealp{keel95, meisenheimer97, werner12}). This is generally agreed to be a result of synchrotron emission arising from the same electron population that produces the radio emission (e.g. \citealp{prieto02, mack09}), thus such a cutoff cannot be present if the observed spectrum between radio and optical frequencies is to be maintained. Unfortunately, the large angular size of 3C452 and 3C223 means that suitable images at optical wavelengths (e.g. from the Hubble Space Telescope) of the hotspot regions are not available due to the relatively small field of view and observations generally being centred on the host galaxy, but if such a disparity between the spectrum of the lobe and hotspot emission is common to FR IIs, it is non-trivial to unify a superposition of high energy cutoffs with the observed power law spectrum of the optical hotspots.
 
Perhaps the most plausible argument for the rapid steepening of the spectrum is due to additional processes modifying the observed hotspot emission. A recent study by \citet{mckean16} who investigated the low frequency turnover in the hotspots study of Cygnus A find that the long assumed low energy cutoff (LEC) alone is unable to explain the observed spectrum when constrained by spatially well resolved LOFAR low frequency observations, concluding that at least some significant contribution from absorption processes must be present.

While Cygnus A is currently the only source for which it has been possible to perform such a detailed analysis of absorption in hotspots due to its large angular size, these processes are able to account for the observed spectral flattening seen at lower resolutions without the need to invoke the narrow range of conditions that would be required for the superposition of high energy cutoffs described above. While no turnover is observed in the hotspots of either 3C452 or 3C223 this is likely due to their relatively small angular size resulting in contamination from lobe emission. This emission may also conceal any minor curvature or break in the intrinsic spectrum of the hotspot, hiding the presence of any stochastic acceleration mechanisms such as those discussed by \citet{prieto02}. Under such conditions, the spectrum may become stretched and flatten at lower energies which, once adiabatic expansion within the hotspot is accounted for, would lead us to observe the portion of the spectrum in the lobes that is not modified by these stochastic processes.

Additional effects may also arise from non-homogeneous conditions within the hotspots, particularly with respect to variations in magnetic field strength or if there is a complex magnetic field topology. As this would lead to the electron population of the hotspots ageing differently than those of the lobes, the synchrotron spectrum may not be linked to that of the underlying electron population in a straight forward manner. We therefore suggest that absorption processes, possibly in combination with stochastic processes and/or non-homogenious conditions, are the most likely cause of the observed disparity. Perhaps most importantly, this suggests that the low frequency spectrum of hotspots is \emph{not} representative of the underlying initial electron energy distribution. Regardless of the cause, if we are to understand the impact on the wider FR II population, then determining whether such a discontinuity exists in the majority of FR IIs is vital going forward.

\subsection{Spectral ages and their robustness}
\label{specagediscussion}

As was noted in Section \ref{specageresults}, all of the models tested are able to provide a good description of the observed spectrum with uncertainties in the region of 10 - 20 per cent for the oldest regions of plasma. Although these uncertainties are much larger than those of studies at GHz wavelengths, this is not surprising as while the low-frequency data used in this paper provide excellent insight in to the low energy electron population and constrain parameters such as the injection index, we see from Equation \ref{specbreak} that for a spectral age of 80 Myr the break frequencies for 3C452 and 3C223 are both $\nu_{t} \approx 1.5$ GHz and so curvature due to ageing is dominated by a single data point. In the context of the current investigation, such uncertainties are more than sufficient to reach our desired scientific aims but emphasises the need for broadband observations well above the break frequency if high precision ages are to be obtained.

While the ages presented in this paper agree well with previous studies for 3C223, the lack of higher frequency observations raises the question as to whether this is the cause of the age discrepancy for 3C452. \citet{nandi10} use two additional observations at 4.91 and 8.35 GHz (not suitable for this study due to their lower resolution) which should probe well past the break frequency in the oldest regions of plasma and provide more robust ages. However, there are a number of additional factors which must be considered when determining which, if either, of these ages is correct. The first, and most straightforward to determine, is that of the injection index and magnetic field strength. The injection index of \citet{nandi10} was derived in a similar manner to the methods used here, but was far less well constrained at low frequencies (which dominates the model dependent determination of the injection index) and used much larger spatial regions which are known to impact upon the derived values (e.g. \citealp{stroe14, kapinska15, harwood17}). Equipartition between the radiating particles and the magnetic field strength is also assumed in this case, resulting in a higher value of $B_{eq} \approx 0.8$ nT but, as was shown by H16 through synchrotron/inverse-Compton fitting, the magnetic field strength of 3C452 is only around half this value. In combination with the steeper injection index, we see from Equation \ref{flux} that this will naturally lead to younger ages. However, fixing these parameters to those of \citet{nandi10} and refitting the JP model we find a maximum age of only around 45 Myr. These variations can therefore only partly account for the difference in age.

An additional factor which was not accounted for in previous studies is that of cross-lobe age variations and the superposition of spectra which can also impact on the derived age of a source. Traditional methods for deriving the spectral age as a function of position (and that used by \citealp{nandi10}) have generally relied on the assumption that, while ages within the lobe increase towards the core, transverse variations are negligible allowing `slices' to be taken across the lobe in order for the required signal to noise level to be reached. While for sources such as 3C223 this is a reasonable (if not perfect) assumption to make, it is evident from Figure \ref{3C452_specage} that this is not the case for 3C452. As the superposition of these spectra do not simply result in an averaged spectrum but a more complex model of emission, this effect is likely to be another significant factor contributing to the observed difference in spectral age compared to previous results. The violation of this assumption appears to be fairly common in FR IIs and has been discussed in detail by H13 and H15 with the results presented here further supporting the need for model fitting to be performed on small spatial scales to reduce these effects if reliable ages are to be determined.

Due to improved constraints on the injection index and magnetic field strength, along with the known superposition of spectra and cross-lobe variations in the lower resolution study, we therefore conclude that the age of 3C452 has been previously underestimated and the values presented within this paper are robust for the given models.

\begin{table*}
\centering
\caption{Source energetics and dynamical ages}
\label{revisedtec}
\begin{tabular}{lcccccccccc}
\hline
\hline
Source&Lobe&Length&Radius&Energy density&Volume&$\epsilon$&$p_{int}$&$kT$&$Q$&$t_{dynamic}$\\
&&(kpc)&(kpc)&(J m$^{-3}$)&(m$^{3}$)&&(Pa)&(keV)&(W)&(Myr)\\
\hline
3C223	&	Northern		&	404		&	67.8		&	$3.90 \times10^{-13}$		&	$1.71 \times10^{65}$		&	--		&	$1.3 \times10^{-13}$	&	--		&	--		&	--	\\
3C223	&	Southern		&	390		&	67.8		&	$3.85 \times10^{-13}$			&	$1.65 \times10^{65}$		&	--		&	$1.3 \times10^{-13}$	&	--		&	--		&	--	\\
3C452	&	Total			&	418		&	33.6		&	$3.16 \times10^{-12}$		&	$1.14 \times10^{65}$		&	0.2		&	$1.1 \times10^{-12}$	&	0.86		&	$7.6 \times10^{37}$		&	152	\\
\hline
\end{tabular}

\vskip 5pt
\begin{minipage}{16.3cm}
`Lobe' gives the region considered in subsequent columns for the target listed in the `Source' column. `Length' and `Radius' list the physical dimensions of the fitted region used to derive the `Volume' and `Energy density'. The term energy density here refers to the sum of the energy density in the radiating particles, the magnetic field, and any non-radiating particles such as protons as used by \citet{harwood16}. `$p_{int}$' lists the internal pressure of the source with `$\epsilon$', `$kT$', and `$Q$' the parameters used in determination of the dynamical age, `$t_{dynamic}$'. Note that dynamical model values are not available for 3C223 as detailed in Section \ref{dynamicalmodel}.
\end{minipage}
\end{table*}

\subsection{Revised energetics}
\label{revisedenergetics}

As noted in Section \ref{dataquality}, the angular size of 3C452 as measured from the 368 MHz VLA image differs from that of the original values of \citet{laing83}. We therefore take this opportunity to remeasure the parameters used by \citet{croston04, croston05} (and subsequently H16 which made a direct comparison) to ensure that the values used to derive the sources volumes, and subsequent total energy content of the lobes, are robust. We find that for both 3C452 and 3C223, even with the additional emission observed due to the low frequencies and increased sensitivity, the volumes are smaller than those found previously. This is most likely due to the lower resolution images used in the original studies and leads to an increased total energy content for both sources, the revised values for which are given in Table \ref{revisedtec}. This does not significantly impact on the findings of H16 as it only acts to further increase the internal pressure of these sources and provides additional support for the conclusion that both sources are at minimum in pressure balance with the external medium.

Uncertainty in the energetics of FR IIs due to their size and, as discussed by H16, low frequency spectrum is perhaps currently a bigger concern to our understanding of the impact these sources have on their environment than any uncertainties in, for example, their intrinsic age (discussed further in Section \ref{dynamicalmodel}). However, upcoming radio surveys such as the LOFAR Two-metre Sky Survey (LOTSS; \citealp{shimwell16}) and the MeerKAT MIGHTEE survey \citep{jarvis12} will provide a large number of well resolved sources over a wide frequency range, and so this problem should be relatively straightforward to resolve based on our new understanding of their spectrum in the near future.

\subsection{Dynamical ages}
\label{dynamicalmodel}

To estimate the dynamical ages of these sources, which extend well beyond the core radii of their host groups, we need a model that can take account of the density profile, rather than assuming a constant external density as in H15 or a power law in density as in \citet{kaiser97}.

We consider a toy model of the expanding radio lobe as a cylinder with length $R$ and cross-sectional area $A = \epsilon R^{2}$, where $\epsilon$ is effectively a measurement of the aspect ratio: $\epsilon =  V / R^{3}$. Then, assuming non-relativistic growth of the source, ram pressure balance requires that \begin{equation}\label{pext1} \rho_{ext} \left(\frac{\rm{d}R}{\rm{d}t}\right)^{2} = p_{int} - p_{ext} \end{equation}

The external pressure and density profiles (functions of $R$) are provided by the modelling of the X-ray observations. Let the one-sided jet power be $Q$ and assume that the jet is light and relativistic so that the jet momentum flux is $Q/c$; this momentum flux is distributed over the front area $A$ of the lobe. Then $p_{int}$ contains a term from this momentum flux and also one from the energy stored in the lobes. We assume here that the lobes are filled with a light, relativistic fluid, so that $p = U/3$, and that, as found in simulations \citep{hardcastle14, english16} about half of the total energy supplied by the jet goes into heating the shocked material between the bow shock and the contact discontinuity (which, as it is necessarily in pressure balance with lobe material at the lobe head, we need not consider further here). Then we have \begin{equation}\label{pext2} \rho_{ext} \left(\frac{\rm{d}R}{\rm{d}t}\right)^{2} = \frac{Q}{c \epsilon R^{2}} + \frac{Qt}{6 \epsilon R^{3}} - p_{ext} \end{equation} where \begin{equation}\label{pext3} \rho_{ext} = \rho_{0} \left[1 + \left(\frac{r}{r_{c}}\right)^{2}\right]^{-3 \beta / 2} \end{equation} and \begin{equation}\label{pext4} p_{ext} = \rho_{ext} k T / \nu \end{equation}

If we further assume that $\epsilon$ is a constant (effectively a self-similar assumption) then the resulting non-linear differential equation for $R$ may be solved numerically for a given $Q$ and external conditions, given some initial condition of the form, e.g. $R = ct$ for $t \ll 1$ Myr (since the equations are singular at $R = 0$, $t = 0$). The simplifying assumption on $\epsilon$ is not really justified, since numerical modelling shows that it varies systematically over the lifetime of a source; however, detailed modelling would have to consider not just the transverse growth of the source but also the behaviour of the swept-up material behind the bow shock, which is beyond what is required for the present paper. When the solution of Equation \ref{pext2} is compared
with the results of numerical simulation of light, relativistic jets we see reasonably good agreement on timescales of a few tens of Myr \citep{english16}. Differences at early times can be attributed partly to the constant $\epsilon$ assumption and partly to the fact that jets in the numerical models are artificially weak at early times due to the coupling of the internal boundary condition to the grid.

The most striking difference is an acceleration of the expansion of the numerical models at late times which is not observed in the solutions to Equation \ref{pext2}. This is because at late times in the models (and in real radio sources) the hot gas pushes the lobes away from the centre of the cluster, so that the lobe volume grows more slowly, and the internal pressure falls less rapidly, than Equation \ref{pext2} would imply. To infer the dynamical age from such a model we measure a source's physical lobe size $R$ and the energy content of a lobe $E = Qt/2$ (again adopting the factor 2 from simulations). We also measure the current value of the lobe aspect ratio parameter $\epsilon = V / R^{3}$. From a family of solutions to Equation \ref{pext2} we can then determine the value of $t$ and $Q$ that most closely satisfies these constraints. The divergence from numerical models once the lobes have moved out from the cluster centre means that these dynamical ages will tend to be overestimates for large age values, but should be reasonable for the two sources we are considering.

Applying the revised values of Table \ref{revisedtec} to the models outlined above we are able to derive a dynamical age for comparison to its spectral counterpart. Unfortunately, for 3C223 a combination of low energy density, the extreme beta model parameters from the XMM fits of \citet{croston04}, and the large core radius of $\sim\,$300 kpc mean that the dynamical models produce unrealistically high dynamical ages on the order of Gyr. This results from the low energy densities forcing the models to low jet powers and a large expansion time through what is, for a significant fraction of its lifetime, a near constant density environment. As was noted by H16, the existing X-ray measurements for 3C223 are highly uncertain and so the beta models and temperature poorly constrained. We therefore conclude that the dynamical model of 3C223 does not currently provide realistic values for the source's age suitable for comparison to the spectral ages derived within this paper.

In the case of 3C452 where we are able to place better constraints on the model parameters, we derive a dynamical age of 152 Myr, almost double that determined from a spectral viewpoint. Such disparities are not uncommon and have long been the topic of discussion (e.g. \citealp{eilek96}; H13) but it has been recently suggested that a simple departure from equipartition between the relativistic particles in the lobes and the magnetic field may be able to resolve such difference in certain cases (e.g. H15, \citealp{turner16}). However, as for 3C452 the magnetic field strength is determined through synchrotron/inverse-Compton model fitting (\citealp{croston04, croston05}; H16) this cannot be the cause of the observed disparity between the derived ages.

One possibility is that models of spectral ageing currently do not accurately represent the intrinsic age of FR IIs, a matter which has been the subject of much debate (e.g. \citealp{alexander87, eilek96a, eilek97, blundell00}). The need for new and updated models which account for our improved knowledge of radio galaxies has been highlighted both in the case of integrated models of spectral ageing which, when tightly constrained, are unable to reproduce an accurate description of radio galaxy spectra \citep{harwood17} and by the success of the Tribble model on well resolved scales (\citealp{tribble93, hardcastle13}; H13, H15) by providing a potentially more realistic description of the magnetic field structure. However, these changes alone are not sufficient to account for the large age differences observed and therefore, if the underestimation of the spectral age is the primary cause of the disparity, additional factors such as the source's evolutionary history (e.g. variance in jet power, environment, magnetic field evolution) must be accounted for within the models if they are to provide a reliable measure of a sources intrinsic age.

Alternatively (or perhaps, additionally), the cause of the disparity may lie in the assumptions made within the dynamical models. In a similar way to ages derived from a spectral view point, these naive models do not fully account for the complex dynamics occurring over the lifetime of a source and model parameters are often poorly constrained. This is highlighted by 3C223 where, although the model presented here accounts for a more realistic scenario than a constant or power law density profile, the dynamical model is unable to determine a robust age. Potentially more realistic dynamical models are currently under development (Hardcastle et al., in preparation) but preliminary results suggest these only serve to further the disparity. Due to the limited number of sources on which such analysis has been performed, both dynamically and using the small scale spectral techniques used within this paper, it is therefore unclear if new dynamical models or simply tighter constraints are needed.

Determining conclusively the intrinsic age of radio galaxies therefore remains an open question. While the difference in spectral ages between studies can be explained in the context of the current models, it emphasises the need for parameters to be well constrained on small scales if reliable ages are to be determined, particularly the injection index and magnetic field strengths. Other considerations, such as older regions of plasma being below the sensitivity limit of the telescope and the mixing of electron populations, mean that the spectral ages may also only represent a lower limit on the intrinsic age of a radio source. In the context of the total energy output over the course of a galaxies lifetime and through all phases of activity, this is further complicated by the need to consider possible recurrent activity in FR IIs. Commonly known as double-doubles, recurrent activity has been observed in a number of radio galaxies (e.g. \citealp{lara99, schoenmakers00, saikia06}) and it has previously been suggested that 3C452 possesses such a morphology. However, a detailed multi-frequency analysis by H16 was unable to find any further evidence of remnant lobes. In either case, the analysis that has been carried out within this paper only considers the current phase of activity.

Ultimately, full numerical simulations of FR IIs that include spectral ageing and are able to account for many more parameters than the simple models presented here are required (English et al., in preparation) and may prove to be the long term solution in determining the dynamical age and improving spectral ageing models. Large samples provided by upcoming surveys will be key to this process providing the physical parameters for these simulations and additional information on how the spectral and/or dynamical models should be updated in order to provide an accurate description of FR IIs.

\section{Conclusions}
\label{conclusions}

In this paper we have presented the first well resolved spectral ageing study of FR IIs at low ($< 300$ MHz) frequencies which we use to study the low energy electron distribution and particle acceleration processes occurring in the two nearby radio galaxies, 3C452 and 3C223, uncovering previously unknown features within their spectrum. Addressing the primary science questions posed in Section \ref{specageintro} we find:

\begin{enumerate}
\item When considered on small spatial scales, the spectrum that describes the low energy electron population remains steeper and is more complex than historically assumed. A rapid spectral steepening is observed in both sources at the boundary between the hotspot and lobe regions suggesting that hotspot spectra are not representative of the initial electron energy distribution. We suggest that this is due to absorption processes and/or non-homogeneous and additional acceleration mechanisms within the hotspots. We caution against using such measurements in determining the injection index of FR IIs.\\
\item The injection index (as derived from the lobe emission) remains steep even when considered at low frequencies and on small spatial scales, consistent with previous findings. The spectral age of our target sources are $77.05_{-8.74} ^{+9.22}$ and $84.95_{-13.83}^{+15.02}$ Myr (Tribble model) for 3C452 and 3C223 respectively.\\
\item Even when a non-equipartition magnetic field strength is assumed, a disparity of a factor of $\sim\,$2 is observed between the age of 3C452 when considered from a spectral and dynamical view point. It remains unclear whether the spectral, dynamical, or both models are the cause of this disparity but upcoming simulations that include spectral ageing may help to resolve this issue.\\
\item The questions raised by these findings have potentially significant implications for our understanding of radio galaxies with respect to their age, energetics, and the underlying processes that drive emission in these powerful radio sources. It is clear that for the two sources investigated, the total energy contained within the lobes is much higher than previously thought but analysis of upcoming large surveys is required to determine the impact on the population as a whole.
\end{enumerate}

\noindent In addition to these primary science questions, we also find:

\begin{itemize}
\item A faint extension not present in higher frequency observations is detected in the northern lobe of 3C223.\\
\item Revised measurements of 3C223 at 368 MHz find its physical size to be 796 kpc, placing it within the giant radio galaxy classification range.\\
\item The lobes of both sources are advancing through the external medium at around one per cent the speed of light.\\
\item Significant cross-lobe variations are again observed, reinforcing the need to perform fitting on small spatial scales if accurate spectral ages are to be derived.
\end{itemize}

In order to explore the observed disparity and particle acceleration processes in hotspots further, sub-arcsecond resolution imaging and the spectrum of hotspots in a larger sample of FR II sources will be presented in a later paper in this series.

\section*{Acknowledgements}
\label{acknowledgements}
We wish to thank the anonymous referee whose constructive suggestions have helped improve this paper. JJH wishes to thank the Netherlands Institute for Radio Astronomy (ASTRON) for a postdoctoral fellowship. This research was partly funded by the European Research Council under the European Union's Seventh Framework Programme (FP/2007-2013)/ERC Advanced Grant RADIOLIFE-320745. MJH and JHC are grateful for support from the Science and Technology Facilities Council under grants ST/M001008/1 and ST/M001326/1. GJW gratefully acknowledges support from The Leverhulme Trust. The Low Frequency Array was designed and constructed by ASTRON (Netherlands Institute for Radio Astronomy), and has facilities in several countries, that are owned by various parties (each with their own funding sources), and that are collectively operated by the International LOFAR Telescope (ILT) foundation under a joint scientific policy. We wish to thank staff of the NRAO Jansky Very Large Array of which this work makes heavy use. The National Radio Astronomy Observatory is a facility of the National Science Foundation operated under cooperative agreement by Associated Universities, Inc. This work has made use of the University of Hertfordshire Science and Technology Research Institute high-performance computing facility. This research has made use of the NASA/IPAC Extragalactic Database (NED), which is operated by the Jet Propulsion Laboratory, California Institute of Technology, under contract with the National Aeronautics and Space Administration.

\bibliographystyle{mnras}
\bibliography{lofar_specage}

\bsp	
\label{lastpage}
\end{document}